# Resonant optical tunnelling in planar three-layer photonic microstructures

Yago Arosa, Alejandro Doval and Raúl de la Fuente

Instituto de materiais da USC - iMATUS, Grupo de Nanomateriais, Fotónica e Materia Branda, Departamento de Física Aplicada, Universidade de Santiago de Compostela, E-15782, Santiago de Compostela, Spain

**Abstract:** We present a unified analytical framework for resonant optical tunnelling in three-layer photonic systems embedded in a transparent dielectric medium. Compact expressions for the system transmission are derived using a generalized Fresnel-coefficient approach, allowing us to distinguish two fundamentally different tunnelling regimes depending on the nature of the waves supported in the core layer. When harmonic waves propagate in the core, resonances follow the conventional Fabry–Perot phase condition. In contrast, when evanescent or damped waves are present, resonant tunnelling arises from an amplitude-matching condition governed by the magnitude of the composite reflection coefficient. This distinction leads to qualitatively different transmission characteristics and dissimilar tunnelling outcomes. For clarity, transparent materials are addressed first (including ideal metals), and the role of absorption is analysed later, revealing the transition from unitary resonant tunnelling to attenuated optical tunnelling. Angular–spectral maps are presented to illustrate the general features of each configuration and demonstrate that resonant tunnelling can occur in structures whose thickness exceeds several wavelengths. The results provide a consistent physical interpretation of resonant tunnelling phenomena across dielectric and metal–dielectric multilayers.

**Introduction**

Optical tunnelling (OT) refers to unexpected light transmission across an optical system composed mainly of high reflection components. Although this concept is relatively modern, since it emerges from the analogy with the quantum tunnel effect [1–4], some configurations in which OT arises have been known for a long time. The most notorious, widely known since the time of Newton, is frustrated total internal reflection (FTIR) [5]. At large incidence angles, a dielectric prism in contact with air totally reflects the transmitted light, while an evanescent tail of the electromagnetic signal crosses the prism-air interface. When a similar prism is approached to the first one, this tail penetrates the second prism and produces a weak transmission of light that increases if the two prisms are brought closer. In more recent times (at the end of the 19th century), Charles Fabry and Alfred Perot invented an interferometer that bears their names [6], consisting of two closely spaced plane-parallel mirrors (e.g., two metallic films). Despite the high reflectivity of the mirrors, the system allows light transmission when a singular or resonant condition is fulfilled. In this second case, in contrast to FTIR, we refer to this phenomenon as resonant OT.

New optical configurations for resonant OT have recently been proposed and studied, inspired by the optical-quantum analogy, and thus leading to possible applications in both fields. The simplest possible configurations correspond to three-layer systems, composed of a thin core layer sandwiched between two outer layers exhibiting high reflectance. Different structures can be distinguished depending on whether the layers are made of dielectric (D) or metallic (M) materials. Namely, DDD [7–9], DMD [7–10] and MDM [11–13] structures have been investigated. More complex systems for resonant OT include both multilayers of low- and high-refractive-index media forming either photonic crystals or one-dimensional metamaterials [14–17] and structured metals [18,19]. Similarly, extraordinary transmission has been observed in a single metallic layer patterned with periodic holes [20,21] or with dielectric spheres [22].

In light of the diversity of configurations in which optical tunnelling appears, most studies treat each structure independently. Fabry–Perot etalons are typically analyzed in terms of phase accumulation, while metal–dielectric stacks are described in terms of coupled surface plasmons. Although mathematically related, these phenomena are rarely placed within a single analytical framework. This fragmentation obscures a fundamental physical question: are resonant optical tunnelling effects governed by different underlying mechanisms unrelated? Or can intrinsically distinct physical regimes be linked by a unified framework?

In this work we demonstrate that resonant optical tunnelling in any three-layer photonic system embedded in a transparent medium reduces to two fundamentally different regimes, determined solely by the character of the waves supported in the core layer: phase-driven OT, when harmonic waves propagate in the core, leading to the conventional Fabry–Perot resonance condition; and amplitude-driven OT, when the core supports evanescent or damped waves, in which case resonance emerges from an amplitude-matching condition governed by the magnitude of the composite reflection coefficient. These two regimes are not mathematically equivalent reformulations of the same interference process but arise from fundamentally different physical constraints on the field inside the core layer.

We derive compact analytical expressions for the global energy transmission coefficients using a generalized Fresnel formalism, applicable to both dielectric and metal–dielectric configurations. This unified treatment reveals that the commonly studied structures are not merely variations of a single interference phenomenon but belong to distinct physical classes. These structures include low–high–low- (LHL) and high–low–high- (HLH) refractive index dielectric configurations, as well as plasmonic geometries involving metals. The main physical distinction is the requirement that the composite reflection coefficients must be greater than one in amplitude-driven resonant tunneling, in contrast to conventional Fabry-Perot systems, in which they are below unity.

By placing dielectric and plasmonic multilayers within a single analytical framework, this work provides a consistent physical interpretation of resonant optical tunnelling and clarifies the conditions under which phase and amplitude mechanisms dominate.

We first analyze transparent systems, including ideal metals, in order to isolate the fundamental mechanisms governing resonant transmission. Absorption is then incorporated and it is seen to turn the ideal unitary resonant tunnelling into attenuated optical tunnelling in realistic plasmonic structures. Angular–spectral transmission maps illustrate the general features of each regime and demonstrate that resonant tunnelling may occur even when the effective tunnelling layer exceeds several wavelengths in thickness.

**Fundamentals**

Our goal is to explore the conditions for resonant OT across different three-layer microsystems, for which we will refer to local transmission maxima as resonances. The permittivity profiles of the structures to be studied here are those depicted in Fig. 1. We have found it appropriate to distinguish dielectric layers with a low refractive index (L) from those with a high refractive index (H).

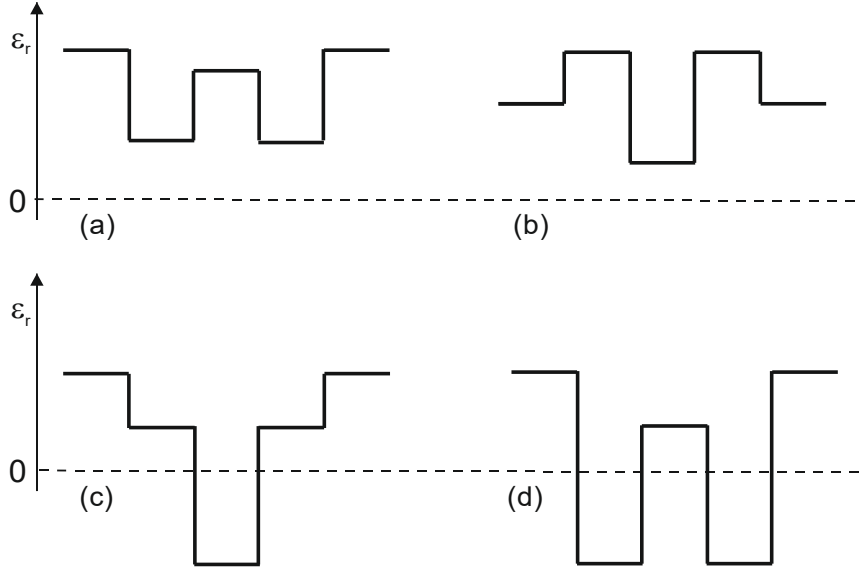

Fig. 1. Some symmetric three-layer configurations embedded in a thick surrounding dielectric that support resonant OT. The vertical axis corresponds to the relative electrical permittivity, which is assumed to be real. We can identify each configuration by its composition. We denote dielectrics with letter L or H, in relation to the low (L) or high (H) value of their refractive index. Ideal metals are labelled as M, and they are characterized by a negative permittivity, i.e. $\varepsilon_r = n^2 = -n''^2$. Using this notation, the different configurations to be studied are: (a) LHL structure (b) HLH (continuous line) structure; (c) LML structure; (d) MLM structure.

Let us start by considering a general $SO_1CO_2S$ structure, composed of a core layer C sandwiched between two high-reflectance outer layers $O_1$ and $O_2$, which are embedded in a thick, transparent dielectric surrounding medium, S. The action of the core layer and the outer layers can be separated, so that the latter are characterized by the following amplitude reflection and transmission coefficients:

$$r_{jmk} = \frac{-r_{mj} + r_{mk}e^{i2k_{m\perp}w_m}}{1 - r_{mj}r_{mk}e^{i2k_{m\perp}w_m}} \qquad t_{jmk} = \frac{t_{jm}t_{mk}e^{ik_{m\perp}w_m}}{1 - r_{mj}r_{mk}e^{i2k_{m\perp}s_m}} \qquad j,k = c,s \qquad m = 1,2$$

(1)

with $w_m$ the width of the outer layer O$_m$, and $k_{m\perp} = 2\pi \sqrt{n_m^2 - n_1^2 \sin^2 \theta}/\lambda$ the component of the wavevector orthogonal to the interfaces inside this layer, with $\lambda$ the wavelength in vacuum and $\theta$ the angle of incidence in the multilayer. There are four coefficients associated with each of these outer layers, namely $r_{jmk}, r_{kmj}, t_{jmk}$ and $t_{kmj}$. For example, $r_{s1c}$ applies to light incident from the surrounding medium S and reflected by layer O$_1$, with the core layer C on the opposite side. Furthermore $r_{mj}$, $r_{jm}$, $t_{jm}$ and $t_{mj}$ are the Fresnel reflection and transmission coefficients for incidence at the interface labelled JM. Besides, round-trip propagation in the core layer of thickness $d$ adds the factor $e^{ik_{c\perp}d}/(1 - r_{c1s}r_{c2s} e^{i2k_{c\perp}d})$ to the field, where $k_{c\perp} = k'_{c\perp} + ik''_{c\perp}$ is the transversal component of the wavevector in the core layer. Taking all together, the transmission amplitude coefficient for the three-layer structure is obtained:

$$t_{s1c2s} = \frac{t_{s1c}\, t_{c2s}\, e^{ik_{c\perp}d}}{1 - r_{c1s}r_{c2s}\, e^{i2k_{c\perp}d}}$$

(2)

The transmittance of the complete SO$_1$CO$_2$S is then straightforwardly calculated as $T = |t_{s1c2s}|^2$. To facilitate the physical interpretation of the subjacent OT process, transparent systems will be addressed first, and the effect of absorption will be incorporated later. In a transparent layer '$j$', the wavevector component in the direction normal to the interfaces can be either real ($k_{j\perp} = k'_{j\perp}$) or imaginary ($k_{j\perp} = ik''_{j\perp}$). In the former case, there are harmonic waves inside the layer, whereas in the latter case there are evanescent waves. In contrast, in the surrounding semi-infinite media S there are always travelling harmonic waves.

**Regime of harmonic waves in the core**

There are two types of resonant OT effects that can be distinguished depending on the type of waves allowed in the core layer. Let us consider harmonic waves first, which corresponds to the case of a Fabry-Perot (F-P) etalon, where the transmittance is well-known to be:

$$T = \frac{T_0}{1 + F \sin^2 \left( k'_{c\perp}d + \frac{\varphi_{c1s} + \varphi_{c2s}}{2} \right)}$$

(3a)

$$T_0 = \frac{(1-R_1)(1-R_2)}{\left(1-\sqrt{R_1 R_2}\right)^2}; \quad F = \frac{4\sqrt{R_1 R_2}}{\left(1-\sqrt{R_1 R_2}\right)^2}$$

(3b)

in which $R_m = |r_{cms}|^2$ is the reflectance of the outer layer $O_m$ for light travelling inside the core layer, and $\varphi_{cms}$ is the phase of the amplitude coefficient $r_{cms}$. $T = T_0$ is the maximum transmittance—and thus its value at resonance— and $F$ is called the finesse, which are both increasing with the reflectance of the outer layers. In particular, for a symmetric system ($O_1 \equiv O_2$) such as those in Fig.1(a), $T_0 = 1$ and $F = 4\,R_1/(1-R_1)^2$.

Provided that the resonances are defined as local maxima of the transmittance, the resonant condition can be obtained by cancelling the sine in Eq. (3):

$$2k'_{c\perp}d + \varphi_{c1s} + \varphi_{c2s} = 2\pi q \quad q \in \mathbb{Z}^+$$

(4)

In this relation the value of d is not limited. However, since we are only interested in optical microstructures, the core layer thickness $d$ will be restricted here to the micron range. The number of resonances depends on the value of $k'_{c\perp}d$, and it is limited to a few units within that range of thicknesses.

As noted, F-P etalons or interferometers constitute the traditional example of an optical system where this type of resonant OT arises. An alternative modern example[7] features a double barrier photonic structure comprising low-refractive index dielectric outer layers, located between high-index core layer and surrounding medium. The case of interest is that of incidence angles $\theta$ from the surrounding larger than the critical angles $\theta_{cr}^{km}$ ($m = 1, 2$, $k = c, s$), defined as a function of the corresponding refractive indices by $sin\theta_{cr}^{km} = n_m/n_k$. In that scenario, waves inside the low-index outer layers are evanescent, with an imaginary wavevector component in the direction normal to the interfaces (let us remember that we are considering harmonic waves inside the core layer). This assures high reflectance for the $SO_mC$ subsystems alone, which is only limited by FTIR. As a consequence, the transmittance of the whole system at resonance and the finesse are also high, until the opaque-layer limit ($R \to 1$) is achieved. This phenomenon can be called resonant FTIR. Amazingly, this resonant OT effect is more prominent for thicknesses of the outer layer where usual FTIR is reduced.

The transmittance of a symmetric system that features resonant FTIR is shown in Fig. 2, as a function of the wavelength and the incidence angle (the F-P etalon is omitted, since it is well known). The core layer and the surrounding media are both made of TiO$_2$, $n(589\ nm) \approx$ 2.61, while the outer layers are SiO$_2$ glass, $n(589\ nm) \approx 1.46$. It corresponds to the permittivity profile depicted in Fig.1a. The results shown correspond to transverse electric (TE) polarized waves, but similar ones are obtained for transverse magnetic (TM) polarization. Two different regions can be visually distinguished within the map, below and above the spectral curve of critical angle, $\theta_{cr}(\lambda)$, for the TiO$_2$ - SiO$_2$ interfaces. On the one hand, below $\theta_{cr}(\lambda)$, there is high transmittance of white light, which presents some typical modulations due to interference effects. On the other hand, above $\theta_{cr}(\lambda)$, the OT resonances arise, exhibiting a broader spectral width, that means low finesse, the closer to the critical angle curve. That bandwidth reduces as the angle of incidence increases, so that resonances show a higher finesse at larger angles. Indeed, the resonances seem to have disappeared for incidence angles greater than 60°, but that is merely an artifact resulting from the limited resolution of the graph.

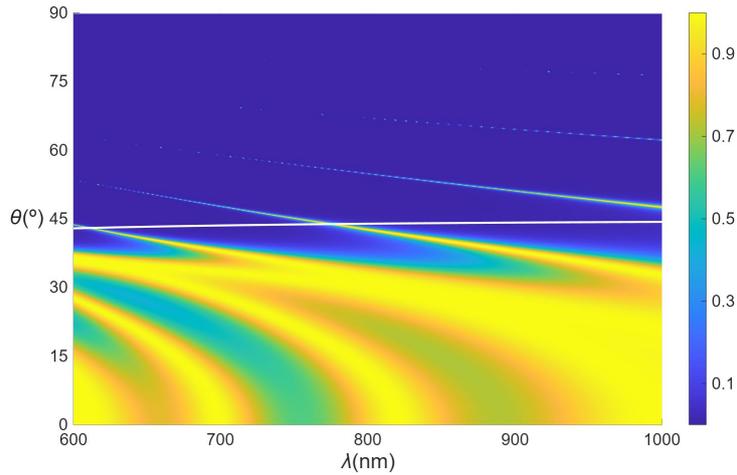

Fig. 2 Transmittance map ($\lambda$, $\theta$) for TE waves at a TiO$_2$- SiO$_2$-TiO2- SiO$_2$-TiO$_2$ structure with core layer thickness $800\ nm$ and $400\ nm$-thick outer layers. The white line corresponds to the spectral curve of the critical angle, $\theta_{cr}(\lambda)$.

**Regime of evanescent waves in the core**

Up to now, we have considered resonant OT effects such that waves inside the core layer are harmonic. Let us now address the other case, in which evanescent waves are allowed inside the core layer. Under these conditions, Eq. (2) leads to the following transmittance:

$$T = \frac{T_0}{1 + F \sinh^2(k''_{c\perp}d - 0.5 \ln|r_{c1s}r_{c2s}|)}$$

$$T_0 = \frac{|t_{s1c}t_{c2s}|^2}{4|r_{c1s}r_{c2s}|\sin^2\left(\frac{\varphi_{c1s}+\varphi_{c2s}}{2}\right)} \quad ; \quad F = \frac{1}{\sin^2\left(\frac{\varphi_{c1s}+\varphi_{c2s}}{2}\right)}$$

(5a)

(5b)

Note that, in contrast to Eq. (3), the maximum transmittance $T_0$ and the finesse $F$ are not written here in terms of the reflectance of the outer layers $R_m = |r_{cms}|^2$ because this concept becomes meaningless when there are evanescent waves in the first medium of the associated amplitude reflection coefficient. The expression for $T_0$ can be simplified if the conservation of flux in transparent systems is considered, which results in $|t_{s1c}t_{c2s}|^2 = 4|r_{c1s}r_{c2s}\sin\varphi_{c1s}\sin\varphi_{c2s}|$, and thus:

$$T_0 = \frac{|\sin\varphi_{c1s}\sin\varphi_{c2s}|}{\sin^2\left(\frac{\varphi_{c1s}+\varphi_{c2s}}{2}\right)}$$

(6)

where the phases of the amplitude reflection coefficients $\varphi_{c1s}$ and $\varphi_{c2s}$ are constrained to the interval $[0,\pi]$, so that $T_0$ meets its maximum value $T_0 = 1$ for $\varphi_{c1s} = \varphi_{c2s}$. Moreover, the denominator of $T$ in Eq. (5) has a minimum when the argument of the hyperbolic sine vanishes, leading to $T = T_0$. As a result, the system will show unitary transmission if these two conditions are satisfied simultaneously:

$$d = \frac{1}{2}\ln|r_{c1s}r_{c2s}|/k''_{c\perp} \equiv D(\lambda,\theta) \tag{7a}$$

$$\varphi_{c1s}(\lambda,\theta) = \varphi_{c2s}(\lambda,\theta) \tag{7b}$$

Condition (7a) means a maximum in transmittance, i.e. a resonance, while condition (7b) assures unitary transmission at resonance. This last condition is always met for symmetric structures, but it can only be fulfilled for some values of $(\lambda,\theta)$ in the case of asymmetric arrangements. Hence, not all solutions of Eq. (7a) can be considered as proper resonances, since the corresponding value of $T_0$ could be small. Conversely, some situations can be considered as resonances even if Eq. (7a) is not satisfied. For example, in a symmetric structure, for a constant $\lambda = \lambda_0$, a resonant branch in the $(d,\theta)$ plane with $T = T_0 = 1$ corresponds to points in which Eq.(7a) is fulfilled until $D(\lambda_0,\theta)$ reaches its maximum value $D_{max} = D(\lambda_0,\theta_m) = d_m$. For $d > d_m$, the resonance persists at the same angle $\theta_m$, but transmission becomes lower as the core layer thickness grows, since that increases the argument of the sinh function in Eq. 5(a). In short, any resonance must approximately satisfy Eq. (7a) and (7b) but can be extended sometimes to situations in which at least one of the two conditions is not verified.

Furthermore, since $d$ and $k_{c\perp}''$ are both positive, the following condition is necessary for the verification of Eq. (7a):

$$G(\lambda,\theta) = |r_{c1s}r_{c2s}| > 1$$

(8)

This condition is precisely the opposite of the usual case with harmonic waves in the core and surroundings in which the coefficients are roots of a reflectance, and thus lower than unity. This condition is related to the existence of a pole in the complex analytical extension of the composite reflection coefficients. In the regime of evanescent waves in the core layer, the transmission and reflection amplitudes have the generic form $t,r \propto [1 - r_{c1s}r_{c2s}e^{-2k_{c\perp}''d}]^{-1}$, and extending $k_{c\perp}''$ to complex values reveals the analytic structure of this denominator. A pole occurs when the denominator vanishes, which corresponds to a self-consistent field in the core sustained by multiple reflections at the outer layers. The inequality $|r_{c1s}r_{c2s}| > 1$ is the necessary condition for such a cancellation to be possible when the core propagation is evanescent, and it therefore delineates the set of multilayer configurations that can support amplitude-driven resonant tunnelling.

Some of the configurations that fulfil eq. (8) correspond to the permittivity profiles shown in Fig. 1(b-d). Only symmetric structures are considered, since they allow unitary transmission. Expressions for $G(\lambda,\theta)$ in each case are given in table 1. Note that metals have been considered ideal, with a negative real dielectric constant.

Table 1. Expressions for $G(\lambda, \theta) = |r_{c1s}r_{c2s}|$ for the three last configurations shown Fig. 1.

| Config. | Conditions | $G(\lambda, \theta) = |r_{c1s}r_{c2s}|$ |
|---|---|---|
| 1(b) TE/TM Waves | $r_{c1s} = r_{c2s}$<br>$\theta > \theta_c = sin(n_c/n_s)$<br>$r_{1s} = \pm\rho_{1s}, \quad \rho_{1s} \leq 1$<br>$r_{1c} = e^{i\varphi_{1c}}$ | $1 - \dfrac{4\rho_{1s}\sin(2k''_{1\perp}w_1)\sin(\varphi_{1c})}{(1-\rho_{1s})^2 + 4\rho_{1s}sin^2(k'_{1\perp}w_1 + \varphi_{1c}/2)}$ |
| 1(c) and 1(d) TM waves | $r_{c1s} = r_{c2s}$<br>$\theta > \theta_c = sin(n_1/n_s)$ 1(c)<br>$\theta > \theta_c = sin(n_c/n_s)$ 1(d)<br>$r_{1c} = \pm\rho_{1c}, \quad \rho_{1c} \geq 1$<br>$r_{1s} = e^{i\varphi_{1s}}$ | $1 + \dfrac{(\rho_{1c}^2 - 1)(1 - e^{-4k''_{1\perp}w_1})}{\left(1 - \rho_{1c}e^{-2k''_{1\perp}w_1}\right)^2 + 4\rho_{1c}e^{-2k''_{1\perp}s_1}sin^2(\varphi_{1s}/2)}$ |

First, Fig. 1b corresponds to a configuration with a low refractive index dielectric core layer. The outer layers are made of a material with a high refractive index (HLH structure), higher than that of the surrounding medium. In this configuration, Eq. (8) is only verified for some continuous values of θ and λ, depending on the signs of the sine functions in the expression of $G(\lambda, \theta)$. For instance, for a constant incidence angle, $\theta_0$, $G(\lambda, \theta_0)$ oscillates periodically, alternating zones with values smaller and greater than 1. The greatest and lowest values of $G(\lambda, \theta)$ are achieved for grazing incidence, for which $\rho_{1s} \to 1$.

An example of an $D(\lambda, \theta)$ map corresponding to this kind of structure is shown in Fig. 3a, together with the transmittance map for the same structure, $SiO_2$ - $TiO_2$ - air - $TiO_2$ - $SiO_2$, in Fig. 3(b-c). Following the shape of $G(\lambda, \theta)$, the function $D(\lambda, \theta)$ is sharply peaked near grazing incidence, with pronounced maxima and minima. If we follow the curve for each maximum or minimum, $|D(\lambda, \theta)|$ decreases as $\theta$ does. However, it increases again near the critical angle, due to the greater value of $1/k''_{c\perp}$.

The transmittance maps in Fig. 3(b) corresponds to those configurations with core layer thicknesses of $d = 150\ nm$. A series of double resonances can be seen near grazing incidence, each of which degenerates into a single resonance for lower incidence angles. These resonance branch shapes are related to those of the $D(\lambda, \theta)$ patterns in Fig. 3a. The merging into a single resonance is accompanied by a reduction in transmittance, since Eq. (7a) is no longer fulfilled. $T$ increases again as $\theta$ approaches $\theta_{cr}$, since $D(\lambda, \theta)$ approaches $d$ again. This map includes transmittance for $\theta < \theta_{cr}$, where resonances are F-P type and verify Eq. (4). A very different behaviour appears in fig. 3(c) where we show transmittance for this kind of structure with $d = 450\ nm$. Whereas the number of resonances above the critical angle is the same, because such number is determined by the product $k''_{m\perp}s_m$, and thus by the thickness of the outer layers, the transmittance at resonances decreases with the thickness $d$ of the core layer. In contrast, the number of resonances of F-P type, i.e. for angles below $\theta_{cr}(\lambda)$, varies. This happens because it depends on the product $k'_{c\perp}d$, and therefore on the thickness of the core layer. However, note that in both cases there is a smooth transition of transmittance at the frontier of the two regimes.

Lastly, we note that a different variant of this resonant OT arises when the dielectric in the core layer of this structure is replaced by a metallic layer[8].

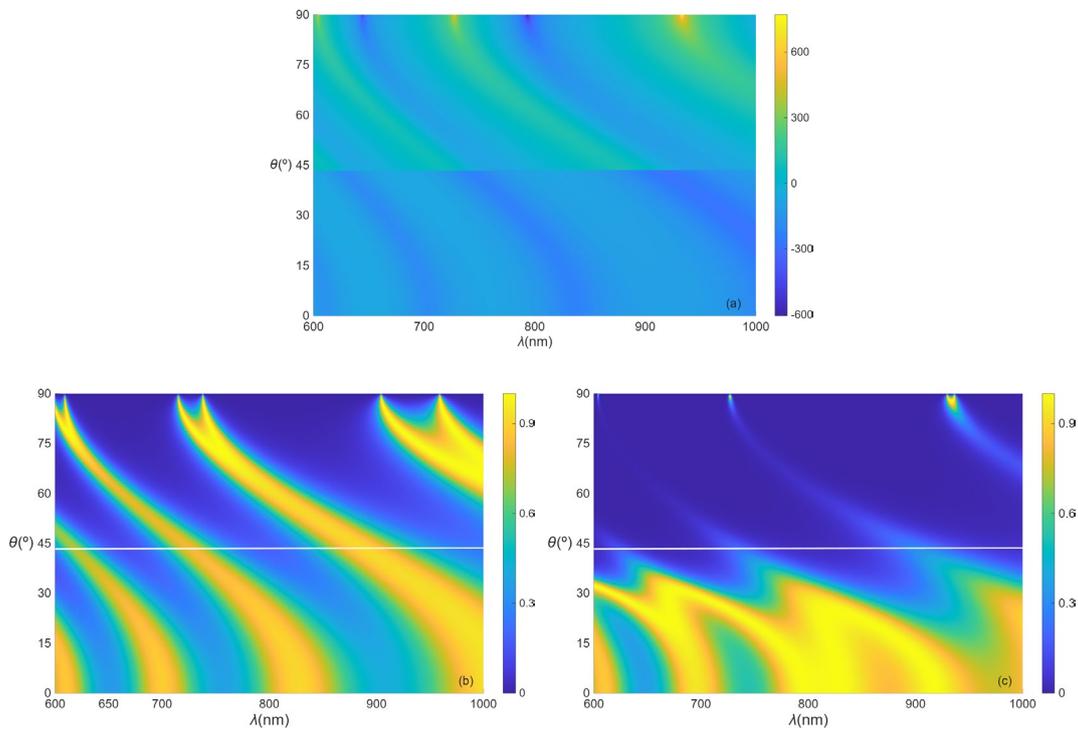

Fig. 3. (a) Map of function $D(\lambda, \theta)$ for TE waves in a $SiO_2$ - $TiO_2$ – Air - $TiO_2$ - $SiO_2$ configuration with $w_m = 1000\ nm$. Transmittance map for this structure with a core layer thickness of $d = 150\ nm$ (b) and $d = 450\ nm$ (c). The white line corresponds to the spectral curve of the critical angle, $\theta_{cr}(\lambda)$. While all the figures are plotted for angles between 0º and 90º, the information in (a) below $\theta_{cr}(\lambda)$ is not relevant

The next cases to be studied are those displayed in Fig. 1c and 1d, which are obtained from one another by interchanging the core and outer layers. The former corresponds to an LML multilayer, where a metallic core layer is sandwiched between two dielectric layers with a refractive index lower than that of the surrounding medium S. The latter, in turn, corresponds an MLM multilayer.

Indeed, the expression for $G(\lambda,\theta)$ is identical in both cases, as indicated in table 1. As listed in this table, $G(\lambda,\theta) > 1$ above the critical angle (for $\theta > \theta_c$) in any case for TM waves, since $\rho_{1c} > 1$. However, Eq. (8) is never satisfied in the case of TE waves, for which $\rho_{1c} < 1$. The resonances in these configurations are related to the generation of coupled surface plasmons (CSP) [13,23]. $G(\lambda,\theta)$ and transmittance maps $(\lambda,\theta)$ for an example of each of the two structures are shown in Fig. 4.

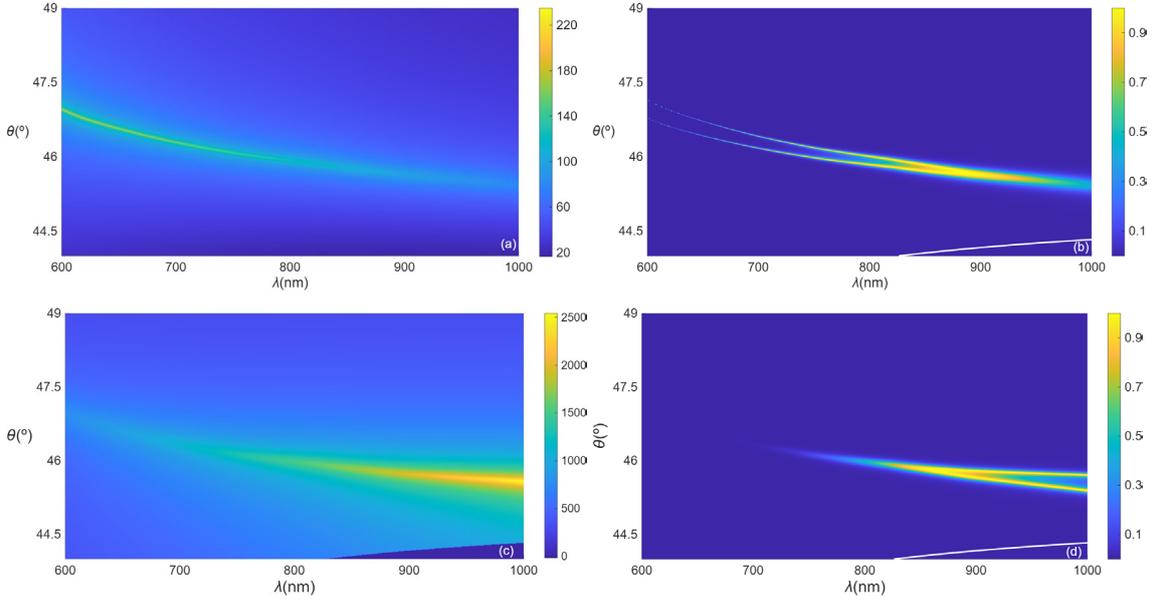

Fig. 4. $G(\lambda,\theta)$ (a) and transmittance (c) for an TiO$_2$-SiO$_2$-Ag*-SiO$_2$-TiO$_2$ structure (Ag* stands for ideal Ag metal, i.e. with a real permittivity) with $w_m = 900\ nm$ and $d = 100\ nm$. $G(\lambda,\theta)$ (b) and transmittance (d) for a TiO$_2$-Ag*- SiO$_2$-Ag*-SiO$_2$ structure with $w_m = 50\ nm$ and $d = 1800\ nm$. The white lines in (c) and (d) correspond to the spectral critical angle curve $\theta_{cr}(\lambda)$.

In both LML and MLM cases represented in Figs. 4(a) and 4(b), $G(\lambda,\theta)$ is observed to have a single branch of high value, beginning at larger incidence angles and shorter wavelengths, and progressing toward smaller incidence angles and longer wavelengths. Given that $k''_{c\perp}$ is a monotonically increasing function of $\theta$, $D(\lambda,\theta)$ will also have a local maximum for each $\lambda$ along a similar curve in the $(\lambda,\theta)$ plane. Let us denote this curve as $D_{max}(\lambda,\theta)$. For wavelengths such that $D_{max} > d$, Eq. (7a) will be satisfied at two points, located on either side of $D_{max}$, leading to two branches of maximum unitary transmission (remember that Eq.

7(b) is always fulfilled for symmetric systems). Otherwise, for other wavelengths $D_{max} < d$ and, in that zone, the two resonances in transmittance degenerate into a single resonance curve. Along this curve, $T$ is now smaller since Eq.(7a) is not fulfilled because $D(\lambda, \theta)$ does not reach $d$ for any resonance angle $\theta$.

The most significant difference regarding transmittance for the two structures lies on the spectral regions where the two resonant branches remain distinct or are merged into a single curve. In LML systems (Fig. 4a and 4c), the maximum of $G(\lambda, \theta)$ is higher for shorter wavelengths and larger incidence angles. This leads to a higher $D_{max}$ and thus to two different resonances within that zone, which become degenerated for longer wavelengths. In contrast, the reverse situation occurs in MLM systems (Fig. 4b and 4d), where the greater maximum of $G(\lambda, \theta)$ and the separated resonances occur for longer $\lambda$ and smaller $\theta$. Another remarkable difference is that MLM systems admit F-P type resonances for $\theta < \theta_{cr}(\lambda)$, while LML systems do not. Indeed, in MLM systems there is a maximum transmittance branch that crosses $\theta_{cr}(\lambda)$, as can be seen in Fig. 4d. This is referred to as the hybrid resonance, since it is of F-P type below $\theta_{cr}(\lambda)$ and plasmonic in nature above it.

**The effect of absorption**

So far, we have considered ideal transparent media. Let us now deal with lossy metals in what follows. Typically, metals in the visible range exhibit a complex dielectric constant, whose real part is negative and greater in magnitude than the imaginary part, i.e. $\varepsilon = \varepsilon' + i\varepsilon''$, where $-\varepsilon' \gg \varepsilon'' > 0$. This affects resonant optical tunnelling in plasmonic configurations (Fig 1 (c-d)), which are the only ones that include metals among the considered arrangements. On the one hand, in this situation, the resonant conditions in Eq. (7), remain valid. However, the position of the resonances changes since $\varepsilon'' \neq 0$ affects the expression of the reflection coefficients $r_{c1s}, r_{c2s}$. However, given that $-\varepsilon' \gg \varepsilon''$, this change is rather small.

On the other hand, the expression for $T_0$ in Eq. 6 must be modified to account for absorption. Limited by that process, unitary transmission is no longer possible. Therefore, we call this phenomenon attenuated optical tunnelling (AOT). Obviously, this imposes limitations on the thickness of the metallic layers within the structure. Maximum transmission values for various examples of the studied structures LMH and HML with varying metallic thicknesses are shown in Fig. 5. For low thickness the transmission is small since there is not sufficient distance to develop the resonance ($d < D_{min}(\lambda, \theta)$). In the case of a metallic core layer, the highest transmission maximum is achieved near 20-25 nm. This peak in maximum transmission is followed by a decay that shows an initial approximately linear trend, which becomes exponential for thicker metal layers. In the case of a dielectric core the highest transmission maximum decreases with wavelength, while the corresponding metallic thickness increases with wavelength. At sufficiently short wavelengths, the chosen core thickness exceeds $D_{max}(\lambda,\theta)$, and, consequently, the transmission diminishes.

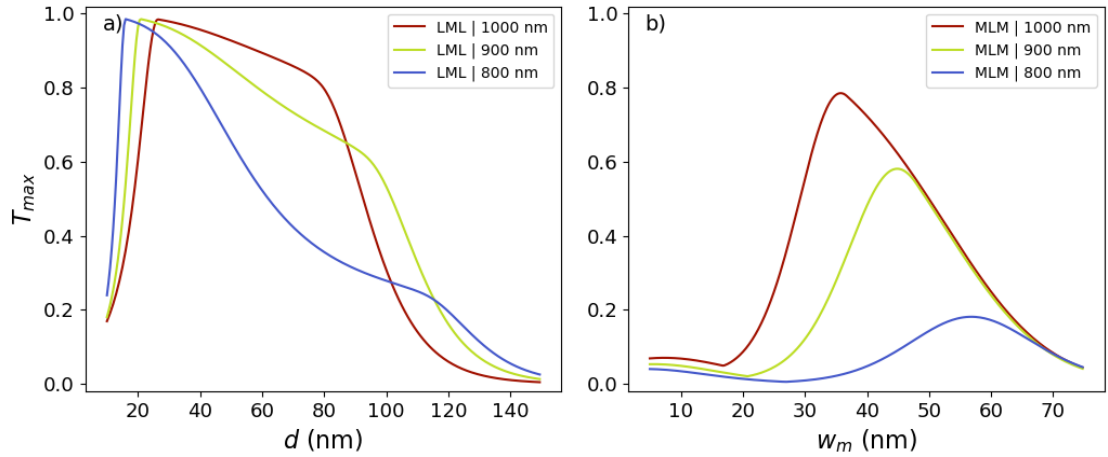

Figure 5. Maximum transmittance value of plasmonic resonances for LML a) (TiO$_2$-SiO$_2$-Ag-SiO$_2$-TiO$_2$) and MLM b) (TiO$_2$-Ag-SiO$_2$-Ag-TiO$_2$) arrangements, as function of the metallic thickness, for 3 different wavelengths. The thickness of the L layer is $w_m$ = 900 nm for LML and d = 1800 nm for MLM structures.

In the case of dielectric layers, absorption does not play a relevant role. However, limitations on the width of the tunneled layers due to the skin depth values must be considered. Let us consider first the LHL multilayer, Fig 1(a). For thicknesses within or below the order of the skin depth, the outer layers exhibit moderate reflectance because of FTIR, resulting in broad, low-contrast resonances. As the width of the outer layers increases, their reflectance approaches one, and the resonances become finer. As shown in [24,25], in the opaque limit ($R_j \to 1, j = 1, 2$), tunneling times are inversely proportional to transmittance ($T_j = 1 - R_j$). Thus, the first outer layer blocks the beam and allows for no tunneling flux. Nevertheless, this is not expected to occur until tunneling layer thickness reach several wavelengths.

On the other hand, resonances in HLH structures are wide and limited to nearly grazing incidence angles. Therefore, we will not discuss this type of structure further here.

Considering the plasmonic structures, those involving metals, the limit on the width of their dielectric layers is related to the maximum of the $D$ function in Eq. 7(a), $D_{max}(\lambda, \theta)$, which is associated as well with the unusually high value of the maximum in $G(\lambda, \theta)$. Considering for example the MLM structure, for $d < D_{max}(\lambda, \theta)$ the system achieves a maximum transmittance of $T_o$, as commented above, only limited by absorption. And for $d > D_{max}(\lambda, \theta)$, the transmittance decreases as thicknesses increase. However, at the same time the maximum of $T_o$ decreases with the of the maximum in $G(\lambda, \theta)$[13]. Therefore, in AOT, there is always a compromise between high transmittance and the high thickness that allows the phenomenon to occur. For example, Fig. 5 shows that high transmittances, above 50 % despite Ag absorption, can be easily achieved for thicknesses of the dielectric layers larger than various wavelengths.

**Summary**

Up to four three-layer configurations for resonant OT, always embedded in a high-index thick media (which can be considered as semi-infinite), have been analysed. Namely, LHL, HLH dielectric structures and LML, MLM structures including metallic layers, have been studied. The first structure differs from the others in both its expression for transmittance and the derived resonance condition (based on phase). This is a consequence of the waves within the core layer being harmonic instead of evanescent or damped, as in the rest of the cases. Thus, resonant optical tunnelling is not a single universal effect, but the manifestation of two different physical regimes governed either by phase coherence or by amplitude compensation. This distinction is not restricted to photonic systems but reflects a general property of wave scattering in layered media.

Regarding the specific configurations studied, the first two (consisting only of dielectric layers) differ notably in the finesse of their resonances, which are very fine for LHL structures and much broader for HLH arrangements. Moreover, the number of resonances is related to the width of the core layer in the former case, and to the width of the outer layers in the latter. In the two last configurations, OT is associated with the generation of CSP resonances, so that there are only two resonances that slowly merge into one as the thickness of the core layer $d$ increases. For sufficiently large thickness, OT vanishes due either to absorption in the metallic layers or to total reflection at the first outer layer. Finally, it has been shown that while absorption limits the width of the metallic layers to tens of nanometres, the thickness of the tunnelled dielectric layers can achieve over-wavelength values.

**Funding**

This research was funded through projects by the Spanish Ministry of Science, Innovation and Universities (PID2024-156552OA-I00), Universidade de Santiago de Compostela (USC 2024-PU031) and Xunta de Galicia (GRC ED431C 2024/06), respectively. Finally, AD thanks the Spanish Ministry of Science, Innovation and Universities for the financial support through FPU21/01302, as well as YA acknowledges Xunta de Galicia for the postdoctoral fellowship ED481D-2024-001.

**Data Availability Statement**

The data that support the findings of this study are available from the corresponding author upon reasonable request.

**Conflicts of interest**

The authors declare that they have no known competing financial interests or personal relationships that could have appeared to influence the work reported in this paper.